\documentclass{JHEP}

\usepackage{epsfig,bbm,bm,amsmath,amssymb}

\def\GN{G_{\rm N}}

\def\be{\begin{equation}}
\def\ee{\end{equation}}
\def\mpr{m_{\rm pr}}
\def\k{{\bf k}}
\def\baray{\begin{eqnarray}}
\def\earay{\end{eqnarray}}
\def\Eq#1{Eq.\ (\ref{#1})}

\def\lsim{\mbox{~{\protect\raisebox{0.4ex}{$<$}}\hspace{-1.1em}
	{\protect\raisebox{-0.6ex}{$\sim$}}~}}
\def\2pi{\left(2\pi\right)}

\def\nott#1{\setbox0=\hbox{$#1$}                
   \dimen0=\wd0                                 
   \setbox1=\hbox{/} \dimen1=\wd1               
   \ifdim\dimen0>\dimen1                        
      \rlap{\hbox to \dimen0{\hfil/\hfil}}      
      #1                                        
   \else                                        
      \rlap{\hbox to \dimen1{\hfil$#1$\hfil}}   
      /                                         
   \fi}                                         %

\title{Constraining the New Aether:  Gravitational Cherenkov Radiation}

\author{Joshua W.\ Elliott, Guy D.\ Moore and Horace Stoica}

\date{\today}

\abstract{
We study the simplest concrete theory for spontaneous Lorentz violation,
the ``New Aether Theory'' of Jacobson and Mattingly, which is a
vector-tensor gravitational
theory with a fixed-modulus condition on the vector field.  We show that
the observation of ultra-high energy cosmic rays (which implies the
absence of energy loss via various Cherenkov type processes)  places
constraints on the parameters of this theory, which are much stronger
than those previously found in the literature and are also stronger
than the constraints generically arising when gravity displays
sub-luminal propagation.
}

\begin{document}


\section{Introduction}

Lorentz invariance is a cornerstone of modern particle physics and of
general relativity.  Exactly how well constrained is violation of
Lorentz invariance?  This question has received increasing experimental
\cite{some_experiments} and theoretical \cite{some_theory} scrutiny
recently (for a nice review see \cite{Mattingly}).  Partly this is
because one should always test the foundations of theories.  Partly it
is also because Lorentz violation is one possible route by which very
high energy (Planck scale) physics might be detectable at the energy
scales actually available experimentally; we can constrain (rule out or
perhaps find evidence for) theories involving the highest energy scales
if those theories happen to predict any deviation from Lorentz
invariance.  The study of Lorentz invariance violation
has also been stimulated by the apparent non-observation of the GZK
cutoff in the cosmic ray flux \cite{Coleman} (for a recent review see
\cite{cosmic_rays}).

Constraints on Lorentz violating physics are already quite severe.  In
fact, if Lorentz invariance is not a symmetry at the fundamental scale,
then it is generically communicated to low energy scales in an
unsuppressed manner \cite{Pospelov,Lorentz_fine_tune}, leading to
$O(10^{-2})$ violations in experiments which instead provide $<10^{-22}$
bounds on its violation.  Similarly, constraints on high dimension
Lorentz violating operators show that they are excluded if they arise
at the Planck scale, even with coefficients suppressed by several orders
of magnitude \cite{Mattingly2,Gagnon}.  (Supersymmetry provides a
potential way around this problem \cite{Pospelov2}.)

Therefore, the study of Lorentz violation has shifted towards
investigating the spontaneous violation of Lorentz invariance, arising
at a scale smaller than the Planck scale.  If this violation occurs in a
way which distinguishes between different species of matter, then the
constraints again become very severe; so one line of work has
concentrated on spontaneous Lorentz violation due to a field coupled
only to gravitation.

Spontaneous Lorentz violation will occur whenever a field, transforming
nontrivially under the Lorentz group, takes on a vacuum expectation
value.  This is the case, for instance, in the vector-tensor theories
originally investigated by Will and Nordtvedt in the early 1970's
\cite{Will_Nordtvedt}.
Recently, a new twist on this idea, called the ``New
Aether Theory,'' has been proposed
(\cite{Jacobson,Jacobson2}, see also \cite{KosteleckySamuel}).  
In this theory, besides the
metric $g_{\mu\nu}$ and ordinary matter fields, there is a new vector
field $S^\mu$, which is constrained, either by a potential term or by a
Lagrange multiplier, to take on a fixed length.  The most general
Lagrangian up to second order in derivatives for such a theory is,
\be
	S = \frac{1}{16\pi \GN}\int d^4x \sqrt{-g}\left(-R-{\mathcal L_V}
	-\lambda (g_{\alpha\beta}S^\alpha S^\beta-v^2)^2
	+ {\mathcal L_M} \right) \, ,
\label{eq:Lag1}
\ee
with ${\mathcal L_M}$ the matter Lagrangian (for instance, of
the Standard Model) and
\be
{\mathcal L_V} =
\frac{c_1}{v^2} \nabla_{\alpha}S^{\beta}\nabla^{\alpha}S_{\beta}
+\frac{c_2}{v^2}\nabla_{\alpha}S^{\alpha}\nabla_{\beta}S^{\beta}
+\frac{c_3}{v^2}\nabla_{\alpha}S^{\beta}\nabla_{\beta}S^{\alpha}
+\frac{c_4}{v^4}S^{\alpha}\nabla_{\alpha}S^{\beta}
	S^{\mu}\nabla_{\mu}S_{\beta} \, ,
\label{eq:Lagrangian}
\ee
which is the most general kinetic term quadratic in derivatives.
(We follow the notational conventions of \cite{Jacobson,Jacobson2} for
the $c_i$, which is opposite that of \cite{Graesser:2005bg}.)
The potential term will ensure that the vector field $S$ will take a
nonzero timelike vacuum expectation value, $S^2 = v^2$ (we use
$[+,-,-,-]$ metric conventions), which then selects a preferred frame,
the frame where the expectation value is purely in the time direction.
In the limit $\lambda \rightarrow \infty$, this is a
strict constraint; at finite but large $\lambda$, the ``radial''
fluctuation is heavy (though we will argue below that the finite
$\lambda$ theory is not viable).

This ``New Aether'' theory is hardly the unique theory for studying the
spontaneous violation of Lorentz symmetry; but it is the simplest, and
its study may teach us about what physical phenomena best constrain such
theories, and what physical problems are encountered in even trying to
formulate them.  Therefore we consider it an interesting problem to
study what the tightest physical limits on this theory are.  Previous
studies have focused on cosmological tests \cite{Carroll}, with bounds
of order $10^{-1}$ on the $c_i$, and on the contribution of $S$ to the
tightly constrained post-Newtonian parameter
$\alpha_2$ \cite{Graesser:2005bg}, which puts a constraint of order $10^{-7}$
on a complicated function of the $c_i$, constraining the $c_i$ to be
$O(10^{-7})$ (except in special parameter ranges, where there can be
cancellations).
However, we expect that a much stronger constraint can be placed on this
theory by calculating how efficiently relativistic particles will
radiate gravitational and $S$-field excitations.  One then uses the
arrival of very high energy cosmic rays to place a bound on how
efficient such energy loss can be, and therefore on the
theory's parameters.  Similar considerations provide the tightest
constraints on subluminal propagation of gravity \cite{Moore:2001bv}.
However, the application to this problem is richer because there are
five propagating modes, corresponding roughly to the two transverse
gravitational modes and two transverse and one longitudinal mode of the
$S$ field \cite{Jacobson2}, and any of these can be emitted by a high
energy cosmic ray.  

The purpose of this paper is to consider such
processes and to use them to set a limit on the coefficients of the New
Aether theory.  As we shall see, the constraints are extremely strong,
and represent the most stringent restrictions on this theory to date.
Presumably this is because, while previous studies
of this theory \cite{Jacobson,Jacobson2,Carroll,Graesser:2005bg} 
have generally been based on considering it
as a classical theory, our treatment takes the theory to be valid
semi-classically (in particle physics language, at tree level).
While Einsteinian gravity is not consistent as a fully quantum theory, it
is consistent at this same semi-classical level, so we consider it
reasonable to ask for the New Aether theory to be interpreted as a
semi-classical, and not just classical, theory.

The outline of the remainder of the paper is as follows.  In Section
\ref{remarks}, we make some remarks about the model, including bounds
arising from causality, and the difficulties encountered above the scale
where radial excitations occur.  In Section \ref{spin2_constraint} we
derive the constraint on the coefficients arising from the emission of
traceless-tensor modes from high energy hadrons; Section
\ref{longit_constraints} obtains a similar constraint when the
``longitudinal $S$ field'' mode is considered, and Section
\ref{long_pair} obtains another constraint from the emission of pairs of
such excitations--a constraint which persists in the limit that all
$c_i$ are taken to zero at fixed ratio.  Finally, we present a
summary of our findings in the discussion section.

\section{Remarks on the New Aether Theory}
\label{remarks}

Before discussing constraints on the New Aether theory arising from a
specific process, we will briefly mention some structural  constraints
on the theory.

To get a taste for the problems at hand, consider the theory in the
absence of the potential term, $\lambda=0$, and without a vacuum
expectation value (using $c_i$ rather than $c_i/v^2$ as coefficients on
the different terms).  
%
%
In this case, the back-reaction on
the metric can be neglected.  Since we are also temporarily not
interested in the case with nonzero vacuum value, we can neglect the
$c_4$ term in \Eq{eq:Lagrangian}.  The resulting effective Lagrangian
is,
\be
{\cal L}_{\rm IR} = -c_1 \partial_\mu S_\nu \partial^\mu S^\nu
	- (c_2{+}c_3) \partial_\mu S_\nu \partial^\nu S^\mu \, ,
\ee
where we replaced covariant with ordinary derivatives, since the metric
is taken to be flat, and integrated by parts to make the $c_2$ term look
like the $c_3$ term.

The point is that, even at the classical level, this Lagrangian yields a
theory with an energy which is not bounded from below,
except in the special case $c_1{+}c_2{+}c_3=0$.  Therefore, as soon as
interactions of any sort are included, we expect runaway behavior.  The
fastest way to see this is to consider  the case where $c_2=c_3=0$ so
only the first term is present; then express it in non-covariant
language as
\be
{\cal L} = -c_1
	\Big( \partial_i S_j \partial_i S_j - \partial_0 S_j \partial_0 S_j
	- \partial_i S_0 \partial_i S_0 + \partial_0 S_0 \partial_0 S_0
	\Big) \, .
\ee
This looks like an ordinary kinetic term for three fields named
$S_1$, $S_2$, and $S_3$, and a wrong-sign kinetic term for a fourth
field named $S_0$.  The associated Hamiltonian has positive energies
associated with gradients and time derivatives of the $S_i$, but
negative energies associated with $S_0$:
\be
H = c_1 \Big( \partial_i S_j \partial_i S_j + \partial_0 S_j \partial_0 S_j
	- \partial_i S_0 \partial_i S_0 - \partial_0 S_0 \partial_0 S_0
	\Big) \, .
\ee
This is the Hamiltonian density derived by finding canonical momenta for each
$S_\mu$ field and applying a Legendre transform, but it does not
coincide with the Hamiltonian density
$T_{00}$ derived by variation with respect to $g^{00}$.  However, the
difference is a total derivative, the total energy remains unbounded
from below, and the theory is unstable.

The contribution to $T_{00}$ arising from the $(c_2{+}c_3)$ term also
leads to an unstable theory, but as is well known, the instabilities
cancel and the Hamiltonian is positive semidefinite for the special
``gauge invariant'' case $(c_1{+}c_2{+}c_3)=0$.  In fact, this is the
reason that, when discussing vector fields in flat-space quantum field
theory, gauge invariance is required in any renormalizable, interacting
theory.

The theory with a vacuum expectation value but without a potential is
also unstable except in the gauge invariant case.  
Classically, as soon as there are excitations in the
gravitational and $S$ field degrees of freedom, the nonlinear
interactions will allow a runaway in which positive energy excitations
combined with the negative energy excitations associated with the
wrong-sign component of $S$ grow without bound.
In the quantum theory, even in vacuum and at tree
level, there will be spontaneous production of a pair of positive and a
pair of negative energy excitations; the rate is unbounded and this
process can be used to rule out the theory,
in analogy to ``phantom'' scalar theories \cite{ClineJeonMoore}.

Adding a potential term $-\lambda (S_\mu S^\mu - v^2)^2$ with finite
$\lambda$ does not rescue the theory.  Expanding $S^\mu = v u^\mu +
\delta S$, the potential introduces a mass term for $\delta S^0$,
\be
{\cal L} \supset -4\lambda v^2 (S_0)^2 \, ,
\ee
but since the kinetic term for $S_0$ has the wrong sign, this actually
makes the field tachyonic.  For instance, if $c_2=0=c_3$, the full
Lagrangian for $S_0$ is
\be
(-4\lambda v^2) S_0^2 +\partial_i S_0 \partial_i S_0
- \partial_0 S_0 \partial_0 S_0 \quad \Rightarrow \quad
\omega^2 = -4\lambda v^2 + k^2 \, .
\ee
Changing the sign of $\lambda$ does not help because it destabilizes the
$S_i$ fields, which then have a negative quartic term.

The only way out is to replace the potential with a strict constraint,
$S_\mu S^\mu = v^2$.  This removes one degree of freedom
(perturbatively, $S_0$) from the model, leaving a positive definite
Hamiltonian.
For this reason, in this paper we will take the New Aether theory to
have the condition $S_\mu S^\mu = v^2$ imposed as a strict constraint.
Of course, this nonlinear sigma model is not renormalizable, which means
it lacks a good interpretation on scales larger than $v$
beyond a semi-classical (tree) level.
However, the same is true of gravity and the scale $m_{\rm pl}$, so we
will not take this as a
fundamental objection to considering this theory (though the lower $v$
must be, the more of an embarrassment it becomes
that the theory cannot be considered at the loop level above this scale).

\TABLE{
\centerline{
\begin{tabular}{|lll|} \hline
Wave mode  &  squared speed $s^2$  &  polarization (when $\hat k=\hat z$)\\
\hline
transverse, traceless metric &
	$1/(1{-}c_1{-}c_3)$  &
	$h_{12}$, $h_{11}=-h_{22}$   \\
transverse Aether &
	$\simeq  \frac{c_1}{c_{1}{+}c_4}$, see \Eq{eq:trans_s2} &
	$h_{i3}\simeq [(c_{1}{+}c_{3})/s]S_i $, $i=1,2$\\
trace &
	$\simeq \frac{c_{1}{+}c_2 {+} c_3}{c_{1}{+}c_4}$, see
	\Eq{eq:trace_s2} &
	$h_{00} = -2 \,\delta \!S_0$,
	$h_{33} \simeq 2\,\delta \!S_0/s^2$ \\ \hline
\end{tabular}
}
\caption{\label{table1} Wave mode speeds and polarizations
in the gauge $h_{0i}=v_{i,i}=0$, from \protect\cite{Jacobson2}.}
}

Jacobson and Mattingly have derived the normal modes of propagation of
the theory, their speeds of propagation, and their admixture between
graviton and $S$ field in a particular gauge \cite{Jacobson2}.  
Their results are
reproduced, with a few approximations, in Table~\ref{table1}.  For
future reference, the complete expressions for the transverse and trace
propagation speeds are,
\be
\label{eq:trans_s2}
s^2_{\rm transverse} = \frac{2c_1-c_1^2 + c_3^2}{2(c_1+c_4)(1-c_1-c_3)}
	\, ,
\ee
and
\be
\label{eq:trace_s2}
s^2_{\rm trace} = \frac{(c_1+c_2+c_3)(2-c_1-c_4)}{(c_1+c_4)
	[2(1+c_2)^2 -(c_1+c_2+c_3)(1+c_1+2 c_2 + c_3)]} \, ,
\ee
while the complete expressions for $h_{ii}$ in the trace mode are
\be
\label{eq:trace_h}
h_{11}=h_{22}=-(c_1+c_4) \delta\!S_0 \, , \qquad
h_{33} = \frac{(1+c_2)(c_1+c_4)}{c_1+c_2+c_3} 2 \delta\!S_0 \, .
\ee

We end this section by mentioning what restrictions can be placed on
the coefficients of this theory by demanding stability, the absence of
ghosts, and causality.  This has been addressed previously 
(see for instance \cite{Lim}, \cite{Graesser:2005bg}),
and we have fairly little to add, except to note how easily the
constraints follow from these speeds of propagation of the normal modes.
As shown in the table, there are three distinct propagating modes with
different dispersion relations.  In the limit where the $c_i$ are small
(the relevant limit, since large values are excluded), these can be
understood as the standard transverse traceless graviton mode, the
transverse $S$ field mode $\delta \vec S \cdot \vec k = 0$, and the
longitudinal $S$ field mode, $\delta \vec S \parallel \vec k$ (in the
frame where the $S$ field condensate is in the time direction, which we
will use throughout).

Gradient instabilities appear when a squared speed of propagation is
negative; from the table, this imposes the constraints,
\be
\frac{c_1}{c_1{+}c_4} \geq 0 \, , \qquad
\frac{c_1{+}c_2{+}c_3}{c_1{+}c_4} \geq 0 \, .
\label{constr1}
\ee
Further, the kinetic $(\partial_0 S_i)^2$ term is wrong sign and the $S$
field excitations are ``ghostlike'' (and ruled out by the arguments of
\cite{ClineJeonMoore}) unless
\be
c_1{+}c_4 \geq 0 \, ,
\label{constr2}
\ee
and therefore $c_1 \geq 0$ and $(c_1{+}c_2{+}c_3)\geq 0$ separately.

We impose in addition the constraint that propagation speeds not be
super-luminal.  After all, we are considering a metric theory, which
should be causal in the usual sense.  This constraint imposes,
\be
\frac{c_1}{c_1{+}c_4} \leq 1 \, , \qquad
\frac{c_1{+}c_2{+}c_3}{c_1{+}c_4} \leq 1 \, , \qquad
c_1{+}c_3 \leq 0
\, .
\label{constr3}
\ee
More precisely, the first two constraints should be derived from
\Eq{eq:trans_s2} and \Eq{eq:trace_s2}.  Lim has argued \cite{Lim} that
these constraints are essential, since their violation could allow
propagation around closed curves.  However they cannot be considered to
be on as firm ground as the other constraints.

The constraints we find will be on the sizes of the $c_i$.  
$S_\mu$ can be re-scaled to force $v^2 = 1$, which has been done in the
previous literature.  In this case, one would
say that we are constraining the kinetic terms of $S$ to be small.
Alternately, we can rescale $S_\mu$ to make the $c_i/v^2$ be order 1,
say, $c_1/v^2 = 1/2$, so the kinetic term for transverse excitations 
is canonically normalized; then the vacuum expectation value
$v^2 = 2c_1$ is small, and we would say we are finding constraints on
the size of the vacuum expectation value of the field $S$.  We prefer
this language, since it makes it more clear that what is being
constrained is the size of the spontaneous Lorentz violation.  However,
previous authors have generally chosen $v^2=1$, and we will continue to
define the $c_i$ to follow their notation.

\section{Constraints from Gravi-Cherenkov Radiation}
\label{spin2_constraint}

\FIGURE{
\centerline{
\begin{picture}(240,100)
    \put(0,0){\begin{picture}(100,100)
	\thicklines
	\put(0,50){\line(1,0){40}}
	\put(40,50){\circle*{4}}
	\put(40,50){\line(2,1){40}}
	\multiput(44,50)(4,-4){8}{\oval(8,8)[bl]}
	\multiput(40,46)(4,-4){8}{\oval(8,8)[tr]}
	\end{picture}}
    \put(120,0){\begin{picture}(120,100)
	\thicklines
	\put(0,50){\line(1,0){40}}
	\put(40,50){\circle*{4}}
	\put(40,50){\line(2,1){40}}
	\multiput(44,50)(4,-4){8}{\oval(8,8)[bl]}
	\multiput(40,46)(4,-4){8}{\oval(8,8)[tr]}
	\put(72,18){\circle*{4}}
	\multiput(76,18)(8,8){4}{\oval(8,8)[tl]}
	\multiput(76,26)(8,8){3}{\oval(8,8)[br]}
	\multiput(76,18)(8,-8){4}{\oval(8,8)[bl]}
	\multiput(76,10)(8,-8){3}{\oval(8,8)[tr]}
	\end{picture}}
\end{picture}}
\caption{\label{fig1}\label{processess}
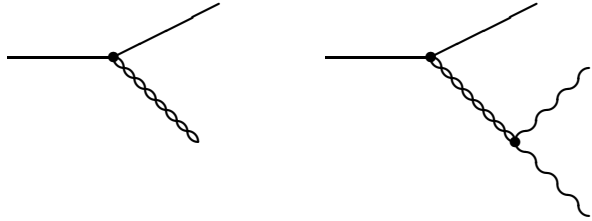
Left:  gravitational Cherenkov radiation.
Right:  emission of a pair of $S$ particles via a virtual graviton.}
}

All constraints we impose on the New Aether theory will arise by
considering the possibility that high energy cosmic rays should undergo
a spontaneous emission process producing one or more excitations of the
coupled gravity-$S$ field modes.  That is, we shall consider diagrams of
the form shown in Fig.~\ref{fig1}, in which the solid line is either a
fermion or a gauge boson, with a dispersion relation we shall take to be
lightlike, and the double-wiggly line is a graviton.  First we consider
the former diagram, with a graviton in the final state.

Normally this process is impossible, because there is no way to conserve
both the energy and the momentum of the final state particles and still
have them on-shell.  This would be true if they both satisfied massive
or light-like dispersion relations.  However, we are investigating a
theory where Lorentz symmetry is broken, and the dispersion relations
are different.  Since in particular the transverse traceless graviton
modes will now have sub-luminal dispersion, with
$k_3^2 = k_0^2/s^2$, $s^2 \leq 1$ the propagation speed in the preferred
frame, these modes
carry more momentum than energy, and it is kinematically allowed to emit
one.  The larger a particle's energy in the preferred frame (assumed to
be the frame defined by the microwave background
radiation%
\footnote%
    {%
    If the preferred frame differs from the frame of the microwave
    background, our bounds actually become stronger.  This is because
    high energy cosmic rays have been observed arriving from many
    directions.  In any frame at a large boost $\gamma \gg 1$ with respect
    to the microwave frame, some of these high energy cosmic rays were
    more energetic than in our frame; our results should be applied to
    these cosmic rays.
    }%
), the higher the emission rate; so the arrival of extremely
high energy cosmic rays sets a limit on how far from light-like the
gravity dispersion relation can be.

The problem of calculating the rate for this process has previously been
solved by Moore and Nelson \cite{Moore:2001bv}, but we will repeat the
derivation here giving somewhat more detail, as well as making a more
careful treatment of the partonic structure of the cosmic ray
hadrons which turns out to slightly strengthen the constraint derived.

Consider the emission of gravitons by a fermionic parton inside of a
hadron.  (Since the energy scale involved in the problem is large, a
partonic discussion is necessary, see for instance \cite{Gagnon}).  In
the gauge used by Jacobson and Mattingly \cite{Jacobson2}, the emitted
excitation is purely a graviton.  The vertex between fermion and
graviton is (see for instance \cite{Lykken})
\be
\frac{-i\sqrt{16\pi \GN}}{8} \left[ \gamma_\mu (P_{1\nu} {+} P_{2\nu})
	+ \gamma_\nu (P_{1\mu} {+} P_{2\mu}) - 2 \eta_{\mu\nu}
	(\nott{P}_1 {+} \nott{P}_2) \right] \, ,
\label{eq:fermion_vertex}
\ee
where $P_1$ and $P_2$ are the 4-momenta of the two external fermion
states.  (Our notation will be that capital letters are 4-momenta, while
lower case letters are the magnitudes of their spatial parts.)
Since the matter Lagrangian is unchanged by including the $S$
field, this vertex is the same; what is different is the dispersion
relation of the final state graviton, which has maximum speed
$s=1/\sqrt{1{-}c_1{-}c_3}\equiv 1-\epsilon$, with
$\epsilon \simeq (-c_1{-}c_3)/2 \ll 1$.  (The assumption $\epsilon \ll
1$ is self-consistent, since we are deriving a constraint on $\epsilon$
which will show that it is order $10^{-15}$.)

Labeling the incoming momentum $P$
and the outgoing graviton and fermion momenta $K$ and $P-K$, we then
have $P^2=0=(P-K)^2$ and $K^2 = -2\epsilon k^2$, where $k=|\vec k|$ is
the magnitude of the graviton spatial momentum in the preferred frame.
It follows that $P\cdot K=-\epsilon k^2$.

The total emission rate of gravitons from this parton is
\be
\Gamma = \frac{1}{2p} \int \frac{d^4 K}{(2\pi)^4} \:2\pi\, \delta[(P{-}K)^2]
	\;2\pi \,\delta[s^{-2} k_0^2{-}|\vec k|^2] \sum |{\cal M}^2| \, ,
\ee
where $\sum |{\cal M}|^2$ is the squared matrix element, summed over
final and averaged over initial spins and polarizations, and we have used the
energy-momentum conserving delta function to perform the integral over
the fermionic final state momentum.  Since the dispersion relation in
the on-shell delta function for the graviton is only simple in the
preferred frame, this is the best frame to evaluate the integral.  The
delta functions can be used to reduce the $K$ integration to two
integrals, over $k$ and $\phi$ a trivial azimuthal angle;
\be
\frac{1}{2p}\int \frac{d^4 K}{(2\pi)^4} \: 2\pi\, \delta[(P{-}K)^2] \;
	2\pi \,\delta[(1{+}2\epsilon) k_0^2{-}k^2]
	= \frac{1}{32\pi^2 p^2}
	\int_0^p dk \int_0^{2\pi}d\phi \, ,
\ee
with $K^2 = -2\epsilon k^2$ and $pk-\vec p\cdot \vec k=\epsilon
k(p-k)$.

The matrix element is easiest to evaluate in the frame where $\vec{k}$
defines the $z$ axis; in this frame, if we choose $\vec p$ to lie in the
$x-z$ plane, $p_x^2 = 2\epsilon p(p-k)$.  We find it most convenient to
compute separately the contribution of the $h_{11} - h_{22}$ and
$h_{12}+h_{21}$ polarizations, with the former giving
\be
\sum |{\cal M}|^2 = \frac{16\pi \GN}{128}(4p_x)^2 \:{\rm Tr}\:
	(\nott{P}{-}\nott{K}) \gamma^x \nott{P}\gamma^x
	= 16\pi \GN \epsilon^2 p(p{-}k)(4p^2{-}4pk{+}k^2) \, ,
\ee
and the latter giving
\be
\sum |{\cal M}|^2 = \frac{16\pi \GN}{128}(4p_x)^2 \:{\rm Tr}\:
	(\nott{P}{-}\nott{K}) \gamma^y \nott{P}\gamma^y
	= 16\pi \GN \epsilon^2 p (p-k)k^2 \, .
\ee
Summing these, and then doing the trivial $\phi$ integral, the
integrated rate is,
\be
\frac{32\pi \GN \epsilon^2}{16\pi p^2} \int_0^p dk p(p{-}k)(2p^2-2pk+k^2)
	= \frac{3}{2} \GN \epsilon^2 p^3 \, .
\ee

Similarly, a gauge boson interacts via the vertex \cite{Lykken}
\be
\frac{-i\sqrt{16\pi \GN}}{2} \left( P_1\cdot P_2 C_{\mu\nu,\rho\sigma}
	+ D_{\mu\nu,\rho\sigma}(P_1,P_2) + \mbox{Gauge dependent}
	\right) \, ,
\ee
where the gauge dependent part vanishes when contracted on external
polarizations, $\mu\nu$ are the graviton polarization indices, $\rho$ is
the Lorentz index for the boson with momentum $P_1$, $\sigma$ is the
Lorentz index for the boson with momentum $P_2$, and
\baray
C_{\mu\nu,\rho\sigma} & = & \eta_{\mu\rho} \eta_{\nu\sigma} +
	\eta_{\mu\sigma} \eta_{\nu\rho} - \eta_{\mu\nu}
	\eta_{\rho\sigma} \, , \\
D_{\mu\nu,\rho\sigma}(P_1,P_2) & = & \eta_{\mu\nu} P_{1\sigma}
	P_{2\rho}-\left[ \eta_{\mu\rho} P_{1\nu} P_{2\rho}
	+ \eta_{\mu\rho} P_{1\sigma} P_{2\nu}
	- \eta_{\rho\sigma} P_{1\mu} P_{2\nu}
	+ (\mu\leftrightarrow \nu) \right] \, .
\earay

The kinematics for graviton emission from a gauge boson are the same as
for a fermion, but the matrix element differs; we find the matrix
element squared, summed over physical final polarizations and averaged
over physical initial polarizations, is
\be
\sum |{\cal M}^2| = 16\pi \GN \epsilon^2 \left[ (2p(p{-}k){+}k^2)^2
	+ k^2 (2p{-}k)^2 \right] \, ,
\ee
leading to an emission rate of
\be
\Gamma = \frac{12}{5} \GN \epsilon^2 p^3 \, .
\ee

The process discussed would cause ultra-high energy cosmic rays to lose
energy.  If the energy loss timescale is shorter than the propagation
time across the galaxy $\sim 10$ Kpc
(the shortest scale on which we believe cosmic
rays must propagate--arguably the highest energy cosmic rays propagate
tens of megaparsecs), then the cosmic rays would not reach us; so the
time scale for the highest energy cosmic rays to propagate without
substantial energy loss must be longer than
10 Kpc $\sim 10^{36} {\rm GeV}^{-1}$.
The transverse momentum ``kick'' a parton experiences in emitting a
graviton, found above, is of order $Q^2 \sim p^2 \epsilon$; this is also
the scale on which parton distribution functions should be evaluated.
We will see in a moment that this is about $Q^2 \sim (3000\:{\rm
GeV})^2$; such a transverse momentum kick is enough to cause a hadron to
fragment, so a single graviton emission should be taken to destroy the
hadron, and the relevant rate is the total rate of graviton emission.
This must be determined by summing over the partons in the hadron.  We
saw that the emission rate scales as $p^3$, so the relevant rate is
\be
\Gamma = \GN \epsilon^2 p^3 \sum_f \int_0^1 dx x^3 f(x) \times
	\left\{ \begin{array}{ll} 3/2 \quad & \mbox{quark}\\
	12/5 \quad & \mbox{gluon} \\ \end{array} \right. \, .
\ee
The right value of $Q^2$ to evaluate the parton distribution functions
$f(x)$ is determined self-consistently from the value of $\epsilon$
saturating the bound;  taking the hadron responsible for the highest
energy cosmic ray to be a proton, and using CTEQ parton distribution
functions \cite{CTEQ}, we find $\sum \int x^2 f(x) \times(3/2\:{\rm or}
\:12/5)\simeq 0.025$.  The result is dominated by the fermion, even
though $(3/2)<(12/5)$, because the $x^3$ weighted PDF is about 6 times
larger for a quark.  With this in mind we will neglect gluon
contributions in the following sections.

Considering that the highest energy cosmic ray which has been detected
arriving at the earth had an energy of
$p\sim 3\times 10^{11}$ GeV \cite{Bird}, and
requiring that it was able to propagate 10 Kpc, we find a bound,
\be
\epsilon \equiv \frac{-c_1-c_3}{2} < 5\times 10^{-16} \, .
\label{bound1}
\ee
This bound is somewhat stronger than the one quoted in
\cite{Moore:2001bv}, because we took a graviton emission to fragment the
hadron, whereas their treatment just summed up the energy losses due to
such emissions.  Note that the emission rate depends on the departure of
the dispersion relation from luminal, $\epsilon$, as $\epsilon^2$.  This
should be compared to ordinary Cherenkov radiation, where the energy
loss rate goes as $\epsilon$.  This is a spin effect.  The polarization
vector/tensor is transverse and only sensitive to the transverse
component of the momentum, which is $O(\epsilon^{1/2})$.  For a spin-1
particle this appears at first order in the amplitude, for a spin-2
particle it appears at second order.  This will be important in what
follows, because the $S$-field polarization states are lower spin, and
will therefore be less suppressed in the $\epsilon \rightarrow 0$ limit.

\section{Emission of $S$ modes}
\label{longit_constraints}

The New Aether theory has five physical propagating modes, rather than
the two familiar from pure general relativity.  This can be simply
understood; three degrees of freedom (the four $S_\mu$ modes, minus the
mode removed by the fixed length condition) have been added.  At small
$c_i$, a gauge can be found in which the propagating modes divide into
the two transverse traceless graviton modes and three $S$ field modes.

The approximate propagation speeds of these three modes are easily read
off from the Lagrangian, \Eq{eq:Lagrangian}, by replacing covariant with
ordinary derivatives, and by setting $S^\mu \nabla_\mu \rightarrow
\partial_0$ in the last term;
\be
{\cal L}_V = S_i \left( \frac{-c_1}{v^2} (k_0^2 - k_l k_l \delta_{ij})
	- \frac{-c_2-c_3}{v^2} k_i k_j + \frac{-c_4}{v^4} k_0^2
	\right) S_j \, .
\ee
Two degenerate modes correspond to $S_i k_i=0$, the transverse Aether
modes, with dispersion relation
$(c_1/v^2 + c_4/v^4)k_0^2 = (c_1/v^2)k^2$.  The remaining mode has
$S_i \parallel k_i$; we shall call it a longitudinal mode.  Its
dispersion relation is
$(c_1/v^2 + c_4/v^4)k_0^2 = ([c_1{+}c_2{+}c_3]/v^2)k^2$.  These
reproduce the small $c$ limits presented in Table \ref{table1}.

If we consider the $v$-scaled $c_i$ to be $O(1)$ coefficients and $v^2
\ll m_{\rm pl}^2$, then it is natural to treat $c_4 \sim c_1^2$.  This
interpretation is natural if the fundamental scale is the Planck scale
but the vacuum expectation value is smaller.  For this reason we will
take the transverse mode to propagate close to the speed of light.
Under this approximation, we need the full expression,
\Eq{eq:trans_s2}, for its speed of propagation $s^2$:
\be
s^2_{\rm transverse} = \frac{ 2c_1-c_1^2+c_3^2}{2(c_1{+}c_4)(1-c_1-c_3)}
	\simeq 1 + \frac{c_1^2 + 2c_1 c_3 + c_3^2 - 2c_4}{2c_1}
	\equiv 1-\epsilon' \, .
\ee
Here $\epsilon' \sim c_i$ is a small parameter indicating the failure of
the transverse mode to move at the speed of light.

When gravity is included, each mode takes on a small admixture of
graviton, and therefore couples to standard model particles and can be
emitted as radiation in processes analogous to the one considered in the
last section.  To compute the emission rate, we must include the leading
order $h_{\mu\nu}$ content of the excitation to find the matrix element,
and we must correctly account for the particle's spectral weight
(correctly normalize the external state) in writing down the external
phase space.

We begin with the transverse $S$ excitation.  The correct on-shell delta
function accounting for the particle's spectral weight (correctly
normalizing the external state) is
\be
2\pi \,\delta \Big( 2(c_1{+}c_4)k_0^2 - 2c_1 |\vec k|^2 \Big) \, ,
\label{eq:state_norm}
\ee
which introduces a $1/c_1$ factor in the rate.  The vertex with a fermion
is \Eq{eq:fermion_vertex} contracted with $h_{13}=h_{31}=(c_1{+}c_3)/s$
(see Table \ref{table1}) for one polarization and
$h_{23}=h_{32}=(c_1{+}c_3)/s$ for the other.
Therefore $|{\cal M}|^2$ introduces two positive
factors of $(c_1{+}c_3)$ and the result is suppressed by $c$ relative to the
result for a graviton.  However, as we will see, the matrix element will
contribute only one, not two, factors of $1-s^2$, so the rate is just as
large as for a spin-2 graviton.

Using the vertex and polarization just presented, choosing coordinates
such that only $k_3$ is nonzero, the matrix element is,
\be
{\cal M} = \frac{i\sqrt{16\pi \GN}(c_1{+}c_3)}{4} \bar u(P{-}K) \left[
	\gamma_1 (2P-K)_3 + \gamma_3 (2P-K)_1\right] u(P)
\ee
for one polarization and $1\rightarrow 2$ for the other polarization.
The spin summed, squared matrix element is
\be
\sum |{\cal M}|^2 = 4\pi \epsilon' \GN (c_1{+}c_3)^2
	\left( 32 p^4 - 32 p^3 k + 42 p^2 k^2 -10 p k^3 + k^4 \right) \, ,
\ee
where $\epsilon'$ is the correction from lightlike dispersion
for the transverse mode, defined above.  The total production rate
from a fermion is obtained by integrating over $k$ and is
\be
\Gamma = \frac{277}{80}
	\frac{(c_1{+}c_3)^2(c_1^2+2c_3 c_1+c_3^2-2c_4)}{2c^2_1} \GN p^3 \, .
\ee
We have not computed the analogous expression for gluons because the
parton distribution function of gluons leads to a smaller contribution.

Using the same integral over parton distributions as in the last section
and applying this bound to the highest energy cosmic ray observed, we
get a constraint,
\be
\frac{(c_1{+}c_3)^2(c_1^2+2c_1 c_3+c_3^2-2c_4)}{2c_1^2}
< 7 \times 10^{-32} \, .
\label{bound2}
\ee
Since the quantity constrained is $\sim c^2$, this constraint is of
comparable strength to the previous one--but has an independent
functional dependence on the $c_i$.

Now we turn to the longitudinal $S$ mode.  The dispersion relation for
this mode, $s^2 \simeq (c_1{+}c_2{+}c_3)/(c_1{+}c_4)$, is the only one which
generically deviates by an $O(1)$ amount from $s^2=1$ even in the limit of
small $c_i$ (small vacuum value $v$ for the $S$ field).  It will
therefore lead to a strong constraint.  But as we will see, the
constraint is even strong when $s^2\simeq 1$, because there is no spin
suppression in the emission of this particle in this limit.

We begin with the phase space.
In the small $c_i$ limit and
in a gauge where the gauge condition is based only on $h_{\mu \nu}$, the
energy-momentum conserving delta function correctly accounting for the
spectral weight (right external state normalization) of the longitudinal
mode is
\be
2\pi \delta \Big( 2(c_1{+}c_4) k_0^2 - 2(c_1{+}c_2{+}c_3)|\vec k|^2 \Big)
	\, ,
\ee
leading to a phase space integration of
\baray
&& \frac{1}{2p} \int \frac{d^4 P_f d^3 kdk_0}{(2\pi)^8}
	2\pi \delta(P_f^2) 2\pi \delta\Big[ 2(c_1{+}c_2{+}c_3)k^2
	- 2(c_1{+}c_4)k_0^2\Big] (2\pi)^4 \delta^4(P{-}P_f{-}K)
	\nonumber \\
&=& \frac{1}{2p} \int \frac{k^2 dk\; d\cos \theta_{pk} \;d\phi}{(2\pi)^3}
	\frac{1}{4|c_1{+}c_2{+}c_3|k}
	2\pi \delta\Big[(p{-}sk)^2{-}p^2{-}k^2{-}2pk\cos\theta_{pk}\Big]
	\nonumber \\
& = & \frac{1}{32\pi|c_1{+}c_2{+}c_3|p^2} \int_0^{\frac{2}{1+s}p}
	dk \, ,
\earay
where $s^2=(c_1{+}c_2{+}c_3)/(c_1{+}c_4)$ is the propagation speed.

However, Jacobson and Mattingly's results \cite{Jacobson2} for the
mixing of the graviton and $S$ field in the state polarization
are in a gauge where $\partial_i S_i=0$.
To use the polarization vectors found there, we must relate the
amplitude of $\delta S_0$ in this gauge to the amplitude of $\delta
S_{\hat k}$
in the above gauge.  The relation turns out to be
\be
\delta S_0\mbox{(Jacobson's gauge)} = s \delta S_{\hat k} 
	\mbox{(above gauge)} \, .
\ee
Therefore the matrix element squared will contain an extra factor of
$s^2$ if the matrix element is computed normalizing the vertex based on
$S_0$, which means $|c_1{+}c_2{+}c_3|$ should be replaced with
$|c_1{+}c_4|$.

The results we have written for the dispersion relation and polarization
tensor in Table \ref{table1} are insufficient to compute the matrix
element because it vanishes at this order;
\be
\bar u(P-K) \left[ \gamma_0 (2P-K)_0 - \frac{1}{s^2} \gamma_3
	(2P-K)_3 \right] u(P)
\propto \bar u(P-K) \left[ \gamma_\mu K^\mu \right] u(P) = 0 \, .
\ee
Therefore we must use the full expressions for the dispersion relations
and the polarization tensor, \Eq{eq:trace_s2} and \Eq{eq:trace_h},
which expanded to the desired order become (choosing
$\hat{k}$ as the $z$ axis)
\be
h_{00} = -s^2 h_{33} - (c_2{+}c_4{-}c_3) S_0
\, , \qquad
h_{11}=h_{22} = -(c_1{+}c_4) S_0 \, .
\ee
These lead to a matrix element squared of
\be
\sum |{\cal M}|^2 = 4\pi \GN (2p-ks)^2 (4p^2-4spk-k^2(1-s^2))
	(c_3{-}c_4)^2 \, ,
\ee
and a total rate of
\be
\Gamma = \frac{2}{15} \GN p^3 \frac{(c_3{-}c_4)^2}{|c_1{+}c_4|}
	\; \frac{6s^3 + 24s^2 + 35s + 20}{(1+s)^4} \, ,
\ee
with $s = \sqrt{(c_1{+}c_2{+}c_3)/(c_1{+}c_4)}$.
If we had not expanded in small $c_i$, then $(c_3{-}c_4)^2$ above would
instead read
$(c_3(2{+}c_4){-}2c_4{+}c_1(c_1{+}c_3{+}c_4))^2/(1{-}c_1{-}c_3)^2$.

The function of $s$ above ranges from 20 to 5.3125 as $s$ goes from 0 to
1; for simplicity we will set it to 5, which will give a conservative
bound.  Since $p_\perp^2$ can be as large as $p^2$ for this process if
$s$ is small, we also use the parton distribution functions evaluated at
$10^{10}$ GeV, which give $\int x^3 f(x)=.0052$ for quarks.  Using the
same constraint on $\Gamma$ as before, we find
\be
\frac{(c_3{-}c_4)^2}{|c_1{+}c_4|} < 1\times 10^{-30} \, .
\label{bound3}
\ee

Curiously, the emission rate which leads to this constraint does {\em
not} vanish in the limit $s\rightarrow 1$ that the emitted mode
approaches lightlike propagation.  This is because its emission, unlike
transverse $S$ and graviton emission, is not spin suppressed.  It can
also be understood as an example of non-decoupling of a longitudinal
mode of a vector field when the mass becomes zero but gauge invariance
is broken.  However, when $(1-s)<10^{-22}$, then it becomes important
that the proton dispersion is not precisely lightlike; so the above
limit is conditional on $(c_2+c_3-c_4)/c_1 > 10^{-22}$.

The emission rate also has a nonzero limit as $s$
goes to zero, $(c_1{+}c_2{+}c_3)\rightarrow 0$, which is the
``gauge invariant'' limit for the kinetic term under which the theory
remains free of tachyons if the
strict fixed-modulus limit on the $S$ field is lifted.
However, if the strict fixed modulus limit is lifted and $\lambda v^2 \lsim
(10^{10} \:{\rm GeV})^2$, then the treatment of the external propagating
states must be reconsidered.  Note that, as $s \rightarrow 0$, the
energy lost by emitting a longitudinal $S$ mode goes to zero.  Naively,
this means that the emission process does not degrade the energy of the
emitter, and is therefore allowed.  The emission process can still be
ruled out, however, because it causes the breakup of the hadron, so no
single final state particle carries as much energy as the initial
hadron, and the highest energy cosmic rays would still be degraded in
energy.%
\footnote%
    {%
    If the highest energy cosmic rays are elementary particles such as
    photons, rather than hadrons, all of our bounds are much stronger,
    since $\int dx x^3 f(x)$ is approximately 1 in this case.  For the
    $s\rightarrow 0$ limit, energy loss occurs whenever the $S$ emission
    is accompanied by bremsstrahlung or pair production.  The inelastic
    cross-section for photon primaries is large enough that the energy
    loss rate still exceeds that for hadrons.
    }

\section{$\psi \rightarrow \psi SS$}
\label{long_pair}

The other process we will consider is the emission of {\em two} $S$
field excitations via an off-shell graviton propagator, the diagram on
the right in Fig.~\ref{fig1}.  This process is suppressed, with respect
to the processes we have just discussed, by an additional power of
$\GN$, so naively it would not be important.  However, the processes we
considered all vanish as the $c_i$ are taken to zero; the graviton
dispersion approaches the light-cone, preventing its emission, and the
direct coupling of matter to $S$ field excitations (via their mixing
with the graviton) is suppressed and also vanishes linearly in $c$.  In
contrast, the coupling of the graviton to matter and to (canonically
normalized) $S$ fields
remains finite as the $c_i$ are taken to zero.  In this limit, the
transverse $S$ field excitation becomes lightlike and the phase space
for its production disappears; but the longitudinal $S$ excitation
remains sub-luminal with $s^2 = (c_1{+}c_2{+}c_3)/(c_1)$ and its
production persists.  Therefore it is possible to set a limit on (ratios
of) the $c_i$ which persists even as the $c_i$  are taken to {\em zero}
(with ratios held fixed).

In studying this process, we will systematically expand in small $c_i$,
which is justified by the results of the previous sections.  We will also
take $c_4 \ll c_{1}$, which we previously argued is reasonable.
Were we not to make this approximation, then transverse $S$ modes could
also be pair produced, and we would find that this process implies that
$c_4/c_1$ is indeed small.  Since the $c_i$ are small, we can
systematically neglect the mixing of graviton and Aether excitations.

We begin by making a crude parametric estimate of the production rate.
At generic values of $s^2$, the squared speed of the longitudinal mode,
the production rate should be of order
\be
\Gamma \sim \frac{\GN^2 p^5}{16\pi} \, ,
\label{eq:est}
\ee
purely on parametric grounds.  Substituting $p \sim 10^{11}$ GeV, this
gives $\Gamma \sim 10^{-23}$ GeV.  This is to be compared with the
propagation time of a typical cosmic ray, $\sim 10^4$ Kpc $\sim
10^{36}/$GeV.  The mean lifetime of a cosmic ray before emitting an $S$
pair is much shorter than the propagation time.  Therefore all parameters
are excluded which do not make $1-s^2 \ll 1$, and we are justified to
expand in this quantity.

\subsection{Kinematics}

We begin with a treatment of the kinematics of the reaction.
The theory has a preferred frame, and it is awkward not to work in this
frame.  Since this is not the rest frame of the emitting particle, and
since the usual Lorentz invariance tricks for treating phase space
cannot be used, the treatment of the kinematics is somewhat more
complicated than would usually be the case for an integrated 3-body
decay rate.
We choose the coordinate axes such that the momentum of the initial
particle is along the z-axis.
The four-momenta of the 3 final-state particles are written as ${\bf
k}_{1}$ and ${\bf k}_{2}$ for the two longitudinal $S$ modes and
${\bf k}_{3}$ for the mode with lightlike dispersion.
All but 5 final state variables are performed by energy-momentum
conserving delta functions.  We choose to leave as integration variables
the following 5 independent variables:
\be
\nonumber
k_{1 z}\,,\;\; k_{3 z}\,,\;\; |{\bf k}_{3\perp}|\,,\;\; \beta\,, \;\;
\text{and} \;\; \phi \, ,
\ee
where $\beta$ is the angle between ${\bf k}_{1\perp}$ and
$-{\bf k}_{3\perp}$. The angle $\phi$ corresponds to an overall
rotation around the z-axis and cannot be fixed by the conservation
of momentum. At the same time, this angle does not appear anywhere in
the matrix elements involved in the emission processes, so the
integral over this independent variable is trivial.

We need expressions for the other variables in the problem in terms of
these 5 independent variables.  However, we do not need absolutely
general expressions, because we can make two approximations.  First,
defining
\be
s \equiv 1-\epsilon \, ,
\ee
we have $\epsilon \ll 1$.  (The estimate, \Eq{eq:est}, shows that
$\epsilon \sim 1$ is already excluded, so we may take $\epsilon \ll 1$
in trying to determine the limit on $\epsilon$.)
This means that, at generic $k_{1z}$ and
$k_{3z}$, the $\bf k$ are all collinear, and we may expand to quadratic
order in the ${\bf k}_{\perp}$.  Second, the graviton propagator will
have a virtuality of $(P-K_3)^2$.  When ${\bf k}_{3\perp}$ is small,
this is close to zero, and the matrix element is enhanced by this nearly
on-shell intermediate state.  For this reason, the dominant region of
the integration is at small ${\bf k}_{3\perp}$, and we may expand in
this quantity being smaller than ${\bf k}_{1\perp}$ and ${\bf
k}_{2\perp}$.

Under the collinear approximation, energy conservation reads
\be
k^{2}_{1\perp}\left(\frac{1}{k_{1z}}+\frac{1}{P{-}k_{1z}{-}k_{3z}}\right)+
k^{2}_{3\perp}\left(\frac{1}{k_{3z}}+\frac{1}{P{-}k_{1z}{-}k_{3z}}\right)
 -2\frac{k_{1\perp}k_{3\perp}\cos\beta}{P{-}k_{1z}{-}k_{3z}}
-2\epsilon(p-k_{3z})=0 \,.
\ee
Further approximating that $\bf{k}_{3\perp}\ll {\bf k}_{1\perp}$, we
find, 
\be
k_{1\perp}^2 = k_{z1} k_{z2} \sqrt{2\epsilon} = k_{2\perp}^2 \, .
\ee

To make the notation in the remainder of the section more compact, we
will define $x=\frac{p-k_{3z}}{p}$, $y=\frac{k_{1z}}{p}$,
$z=\frac{k_{3\perp}}{p}$.
$x$, $y$, and $z$ are independent
integration variables, with $x$ in the range $[0,1]$, $y$ in the range
$[0,x]$, and $z$ ranging up to of order $\sqrt{\epsilon}$, but as
discussed, the rate
will be dominated by the region where $z\ll \sqrt{\epsilon}$, so the
upper bound of this integration will not be important.
The $\beta$ integral runs over the range $[0,2\pi]$ for small $z$, and
at leading order in small $z$ the integrand will not depend on $\beta$,
so this integral may also be conducted trivially.

With these simplifications, the phase space becomes
\baray
&&\frac{1}{2} \: \frac{1}{2p} \int \frac{d^3 k_1\, d^3 k_2 \, d^3 k_3}
	{(2\pi)^9 8k_1 k_2 k_3} (2\pi)^4 \delta(k_{1z}+k_{2z}+k_{3z})
	\delta(p{-}k_1^0{-}k_2^0{-}k_3^0)
	\delta^2(\k_{1\perp}{+}\k_{2\perp}{+}\k_{3\perp})
\nonumber \\ 
& \simeq & \frac{1}{2^{10} \pi^5 p}\int \frac{dk_1 dk_3}{k_1 k_2 k_3}\,
	d\phi \: d\beta \: k_{1\perp} d k_{1\perp} \: k_{3\perp} d k_{3\perp} 
	\delta\left(\frac{k_{1\perp}^2}{2}
	\left[\frac{1}{k_1}+\frac{1}{k_2} \right]
	-2\epsilon(p-k_3)\right)
\nonumber \\
& = & \frac{p}{2^8 \pi^3} \int_0^1 dx \int_0^x dy \int dz
	\frac{z}{x(1-x)} \, ,
\label{eq:phase_space}
\earay
where the small $k_{3\perp}$ approximation was made in passing from the
first to second expressions.
The leading factor of 1/2 in the first expression is the final state
symmetry factor.

\subsection{Matrix Element}

The vertex coupling the graviton to two $S$ field excitations can be
found by expanding \Eq{eq:Lagrangian} to linear order in
$h_{\mu\nu}$.  As discussed above, we neglect $c_4$.
Writing the vertex as
$h_{\mu\nu}(Q)S_{\alpha}(K_1)S_{\beta}(K_2)V^{\mu\nu\alpha\beta}$,
we find
\baray
\label{vertex}
\frac{V^{\mu\nu\alpha\beta}}{\sqrt{16\pi\GN}} & \!=\! & 
-c_1 \left[ \eta^{\alpha(\mu}\eta^{\nu)\beta}
	\left(2K_1\cdot K_2 + Q^2\right)
	 +2\eta^{\alpha\beta}K_1^{(\mu}K_2^{\nu)}
	 +2(K_1{-}K_2)^{(\mu} \eta^{\nu)[\alpha}\, Q^{\beta]} \right]
	\nonumber  \\ &\!\!\!&
- c_2 \left[ 2K^\alpha_{1}(K_2+Q)^{(\mu} \eta^{\nu)\beta}
	- K_1^\alpha Q^\beta \eta^{\mu\nu}
	+2K^\beta_{2}(K_1+Q)^{(\mu} \eta^{\nu)\alpha}
	- K_2^\beta Q^\alpha \eta^{\mu\nu} \right]
	\nonumber  \\ &\!\!\!&
- c_3 \left[ 2 K_1^\beta K_2^{(\mu} \eta^{\nu) \alpha}
	+ 2 K_2^\alpha K_1^{(\mu} \eta^{\nu) \beta}
	+ Q^2 \eta^{\alpha(\mu} \eta^{\nu)\beta}
	 +2(K_1{-}K_2)^{(\mu} \eta^{\nu)[\alpha}\, Q^{\beta]} \right]
	\, .
\earay
This is linear in the $c_i$ and so the matrix element squared is
quadratic in the $c_i$; however, there is a $1/(2c_1)$ factor associated
with each $S$ field external state, see \Eq{eq:state_norm}, so the rate
is zero order in the $c_i$.
Alternatively, if we canonically normalize the
$S$ field from the start, each term in \Eq{vertex} should be multiplied
by $1/(2c_1)$.

In calculating the matrix element we must be somewhat careful, since the
final state phase space remains non-vanishing even as the graviton
becomes lightlike.  Since the matrix element will be singular at this
point due to the graviton propagator, one must treat the kinematics of
the very small $k_{3\perp}$ region carefully, in particular including
the nonzero proton mass in computing $Q^2$, how far off-shell the
graviton is.  (Because of this near-singularity, the dominant $|Q^2|$ will
be $\lsim \mpr^2$ the proton mass squared.  The process is
therefore not probing the structure of the proton, and we can consider
it as a single fermionic object.)
This gives corrections of the form $\mpr^2/p^2$ to the
virtuality of the graviton.

We again use the expression for the graviton-matter coupling from
\cite{Lykken} and take the graviton propagator in the usual Dedonder
gauge. The longitudinal polarization vectors are adequately
approximated by
\be
\vec{\epsilon}_1 \simeq \vec{\epsilon}_2 \simeq \hat{z} \, ,
\ee
good at leading order in $\epsilon^{1/2}$.  Therefore,
\baray
\vec{\epsilon}_1 \cdot \vec{p}&=&\vec{\epsilon}_2 \cdot \vec{p}\simeq p,\qquad
\nonumber\\
\vec{\epsilon}_1 \cdot \vec{k}_3&=&\vec{\epsilon}_2 \cdot \vec{k}_3
	\simeq p(1-x).
\earay

Because of the near-singularity in the graviton propagator, the
cross-section is dominated by the kinematic regime $Q^2 \sim \mpr^2 \ll
\epsilon p_0^2 \sim K_1\cdot K_2 \sim P \cdot K_1$.  In this regime most
of the terms in \Eq{vertex} are subdominant; only the terms involving
$c_1 K_1 \cdot K_2$, $c_2 (K_1^\alpha K_2^\mu + K_2^\beta K_1^\mu)$,
and $c_3 (K_1^\alpha K_2^\mu + K_2^\beta K_1^\mu)$ contribute at order
$\epsilon p_0^2$.  Furthermore, since we are specializing to the regime
$\epsilon \ll 1$, we must have $|c_2{+}c_3| \ll c_1$; otherwise the
propagation speed is far from luminal, which we have already excluded.
Therefore the $c_2$ and $c_3$ contributions to the matrix element cancel
in the leading kinematic regime and we are left with only the $c_1$
contribution.  This rather substantially simplifies the calculation.
The matrix element, including the final state normalization factor
mentioned above, reads
\baray
\sum |{\cal M}|^2 = \frac{(16\pi G_N)^2 }{4 Q^2 Q^2}
		\left(K_1\cdot K_2\right)^2
		(\hat{z}\cdot \vec{p})(\hat{z}\cdot \vec{k}_3)
		\left(\hat{z}\cdot \vec{p}+\hat{z}\cdot\vec{k}_3\right)^2.
\earay
In terms of the variables defined in the preceding section this is
\baray
\sum |{\cal M}|^2 = \varepsilon^2 (16\pi \GN)^2 p^4 \;
	\frac{(1{-}x)^3\, (2{-}x)^2 \, [y^2+(x{-}y)^2]^2}{4\left(z^2+
	\frac{\mpr^2}{p^2}x^2\right)^2}\, ,
\earay
which is to be substituted into \Eq{eq:phase_space}.  The $z$ integral
gives,
\be
\int \frac{z dz}{\left(z^2+\frac{\mpr^2}{p^2}x^2\right)^2} = 
	\frac{p^2}{2 \mpr^2 x^2} \, ,
\ee
illustrating that the integral is dominated by small $z$ as claimed.
The remaining $x$ and $y$ integrations are straightforward, and give a
total rate of
\be
\Gamma = \frac{\epsilon^2 p^7 \GN^2}{225\pi \mpr^2} \, .
\ee

Unlike the processes in the previous two sections, in this process the
transverse momentum ``kick'' to the proton is 
relatively small, so the proton typically does not break up, and we should
compute the rate of energy loss rather than the frequency of emission.
This is done by adding a factor $xp$ inside the integral, leading to an
energy loss per time of
\be
\frac{dp}{dt} = \frac{29 \epsilon^2 p^8 \GN^2}{14400\pi \mpr^2} \, .
\ee
Assuming that the energy loss should not exceed the total energy of the
highest energy cosmic ray observed over a 10 Kpc distance, and again
using $p=3\times 10^{11}$ GeV \cite{Bird}, this leads to
a bound on $\epsilon$ of
\be
\epsilon \equiv \frac{c_4-c_2-c_3}{c_1} < 3\times 10^{-19} \, .
\label{bound4}
\ee
We see that the approximation $(\mpr^2/p^2)\ll \epsilon$ is
self-consistent.

\section{Discussion and Conclusions}

Spontaneous Lorentz violation requires that a field, transforming
nontrivially under the Lorentz group, take on a vacuum expectation
value.  The simplest available field is a vector field, and the most
general theory where such a field takes on a vacuum value, which is
stable and free of ghosts, is the New Aether theory, with a Lagrangian
given in \Eq{eq:Lag1}, \Eq{eq:Lagrangian}.
This theory is ill behaved unless the
restriction on the modulus $S_\mu S^\mu$ is imposed as a constraint, not
through a potential.  This renders the theory non-renormalizable and
valid only semi-classically.

The most notable feature of the theory at the semi-classical level is
that there are 5 propagating modes---two graviton, two transverse $S$,
and one longitudinal $S$---which are massless but have sub-luminal
propagation.  As a result, high energy particles moving close to the
speed of light can radiate them, in analogy with the Cherenkov process.
This makes high energy particles lose energy--the higher energy the
faster the energy loss.  Since very high energy cosmic rays are known to
travel astronomical distances, this places constraints on the theory.
Namely, it cannot have parameters causing energy loss which would
degrade those high energy cosmic rays.
Since there are several modes with different dispersion, there are
several constraints.  In particular, we have found,
\baray
-c_1-c_3 & < & 1\times 10^{-15} \qquad \mbox{\Eq{bound1}}
	\\
\frac{(c_1{+}c_3)^2 (c_1^2+2c_1 c_3 +c_3^2- 2c_4)}{c_1^2} & < &
	1.4\times 10^{-31} \hspace{1.27em} \mbox{\Eq{bound2}}
	\\
\frac{(c_3 {-} c_4)^2}{|c_1{+}c_4|} & < & 1\times 10^{-30}
	\qquad \mbox{\Eq{bound3}}
	\\
\frac{c_4-c_2-c_3}{c_1} & < & 3\times 10^{-19}
	\qquad \mbox{\Eq{bound4}}
\earay
The next-to-last bound is conditional on 
$(c_4-c_2-c_3)/c_1 > 10^{-22}$.  In every case the opposite side of the
expression is bounded by 0.  This is because no dispersion relation can
be super-luminal in flat space.

Combining these bounds with \Eq{constr1}--\Eq{constr3},
the $c_i$ are constrained such that each must be $|c_i|<10^{-15}$,
except for a special case in which each
dispersion relation in Table~\ref{table1} is almost exactly luminal.
Using the full expressions, \Eq{eq:trans_s2} and \Eq{eq:trace_s2} for
the propagation speeds, this special case occurs when the $c_i$ satisfy
\be
c_3 = -c_1 \, , \qquad c_4 = 0 \, , \qquad
c_2 = \frac{c_1}{1-2c_1} \, .
\ee
We have checked that these same conditions cause $\alpha_2$ as
determined in \cite{Graesser:2005bg} to vanish (assuming the result
quoted there is leading order in the $c_i$).

Physically, this means that Cherenkov processes rule out the New Aether
theory unless either
\begin{enumerate}
\item
parameters are just right such that all three propagating modes
(graviton and two $S$ field) are light-like to extremely high precision,
a 1-dimensional subspace of the 4-dimensional parameter space of the
model, or
\item
canonically normalizing the field $S$,
the vacuum expectation value $v$ of $S$ satisfies
$v < 3\times 10^{-8}m_{\rm pl} \sim 5 \times 10^{11}$ GeV, or
\item
the constraints, \Eq{constr3}, are not correct; there are modes which
propagate super-luminally,
despite the theory being a generally covariant metric theory of
gravity.  (We earlier discounted this possibility because we are
considering a metric theory where the violation of Lorentz invariance is
supposed to be spontaneous.  However, the necessity of the constraint
against super-luminal propagation is controversial, so we acknowledge
this logical possibility.)
\end{enumerate}
The constraints are not fatal to the theory, but they mean that either
the theory does not predict Lorentz violating propagation speeds, or the
scale of the spontaneous Lorentz violation must be much smaller than the
Planck scale.

\medskip
{\bf Acknowledgements}
\medskip

\noindent
This work was supported in part by the National Sciences and Engineering
Research Council of Canada, and by le Fonds Nature et Technologies du
Qu\'ebec.

\end{document}